\font\twelvebf=cmbx12
\font\ninerm=cmr9
\nopagenumbers
\magnification =\magstep 1
\overfullrule=0pt
\baselineskip=18pt
\line{\hfil PSU/TH/196}
\line{\hfil March 1998}
\vskip .8in
\centerline{\twelvebf  On the Hagedorn Transition and Collective Dynamics of
D0-branes}
\vskip .5in
\centerline{\ninerm S.CHAUDHURI and D.MINIC}
\vskip .1in
\centerline{Physics Department}
\centerline{Pennsylvania State University}
\centerline{University Park, PA 16802}
\vskip .1in
\centerline {shyamoli@phys.psu.edu,
minic@phys.psu.edu}

\vskip 1in
\baselineskip=16pt
\centerline{\bf Abstract}
\vskip .1in

Banks, Fischler, Klebanov and Susskind have proposed a model for black hole 
thermodynamics based on the principle that the entropy is of order the number 
of particles at the phase transition point in a Boltzmann gas of D0-branes. 
We show that the deviations from Boltzmann scaling found in $d$$<$$6$ noncompact 
spatial dimensions have a simple explanation in the analysis of self-gravitating 
random walks due to Horowitz and Polchinski. In the special case of $d$$=$$4$ 
we find evidence for the onset of a phase transition in the Boltzmann gas analogous 
to the well-known Hagedorn transition in a gas of free strings. Our result relies 
on an estimate of the asymptotic density of states in a dilute gas of D0-branes. 

\vfill\eject

\footline={\hss\tenrm\folio\hss}

\magnification =\magstep 1
\overfullrule=0pt
\baselineskip=22pt
\pageno=1

Matrix theory [1] has motivated a qualitative description of the physics of 
Schwarzschild black holes in $d$$>$$4$ noncompact spatial dimensions in terms 
of the thermodynamic properties of a Boltzmann gas of D0branes, interacting 
via long range gravitational forces [2-9]. According to this proposal neutral 
black holes may be understood as bound states of the partons of Matrix theory. 
The analysis is entirely within the mean field approximation to the collective
dynamics of D0branes, where the D0-branes are {\it distinguishable} Boltzmann 
particles [3,13-15]. All basic qualitative features of the thermodynamics of 
Schwarzschild black holes are derived using simple scaling arguments.

Let us summarize these scaling relations. In [2,3] the Bekenstein-Hawking 
relation for a black hole in $D$$=$$d+1$ spacetime dimensions was deduced from a mean 
field analysis of the collective dynamics of $N$ Matrix theory partons in the 
limit of low velocities $v$ and large relative separations $r$. The mean field 
Lagrangian of N D0branes, with longitudinal momentum $P=N/R$, is
$$
L  = N {v^{2} \over 2R} + N^{2}{G_{D} \over R^{3}} 
{v^{4} \over r^{D-4}} \quad . \eqno(1)
$$ 
Application of the Virial Theorem, and the Heisenberg uncertainty principle,
$v \sim {R \over R_S}$, to a purported bound state of N D0branes of transverse 
size, $r=R_S$, gives the scaling $ N \sim R_S^{D-2}/G_{D} $. If we assume the 
simple rule of thumb $S \sim N$, we get the Bekenstein Hawking scaling relation,
$$ S \sim R_S^{D-2}/G_{D} \quad , \eqno(2) 
$$ 
which describes the thermodynamics of a $d+1$-dimensional black hole with 
Schwarzschild radius $R_S$ and entropy $S$. The Schwarzschild radius can then be 
shown to scale with the black hole mass, $M$, according to the relation 
$$ R_S \sim (G_{D} M)^{{1 \over {D-3}}} \quad , \eqno(3)
$$
upon using the basic light cone kinematical relation, $E \sim M^{2}R/N $. Note
that, because of the form of the velocity dependent interaction of slow-moving 
D0branes, the manipulations that led to the Bekenstein Hawking identity are 
ill-defined in the important case of $d=4$ noncompact spatial dimensions. In fact, 
from the expression 
$$
S \sim G_D^{{1\over{D-3}}} \left ({{NE}\over{R}} \right )^{{{D-2}\over{2(D-3)}} }  
 \quad , \eqno(4) 
$$ 
we see that it is not possible to match to the scaling relation characteristic of 
a {\it field} theory in $d$ spatial dimensions $S \sim E^{{p}\over{p+1}}$, with 
integer $p$, in the cases $d$$>$$5$ [3]. Furthermore, the assumption $N \sim S$
gives a specific heat that is either infinite or negative in the cases $d$$<$$6$ 
implying a breakdown of the Boltzmann gas model.

We will show in this paper that these facts have a simple explanation in the 
behavior of self-gravitating random walks as have already been analyzed by 
Horowitz and Polchinski [10]. We provide an independent estimate for the 
asymptotic density 
of states of an ensemble of D0branes in four noncompact spatial 
dimensions that is direct evidence of a Hagedorn like transition 
in a gas of free strings. As an aside, we note in passing that the term {\it 
Hagedorn transition}, as opposed to the Hagedorn limiting temperature, was first 
used in the context of phase transitions in the (MIT) bag gas model [31]. 
Not surprisingly, given the analysis of Horowitz and 
Polchinski, in the special case of four noncompact dimensions the effect of 
interactions becomes non-negligible
{\it precisely} at the phase transition point into a black hole. We will 
find that, at the phase transition point, $S \sim {\sqrt{N}}$, in agreement
with the intuition that the correct setting for the $d$$=$$4$ case is the 
regime $N $$>$$S$ [3].

Let us briefly summarize the evidence for a phase transition at $S \sim N$ in a  
Boltzmann gas of D0branes. These arguments do not take into account the influence 
of interactions and are independent of spacetime dimensionality. It has been pointed 
out in [6] that the $S \sim N$ rule of thumb implies $P \sim S/R$, which is believed 
to hold at the transition point at which longitudinally stretched black strings become 
unstable to the formation of black holes. Further discussion of the validity of the scaling 
relation $S \sim N$ from the viewpoint of the thermodynamic properties of a 
Boltzmann gas of distinguishable particles was also given in [3] and in [13-15].

First, we note that the Boltzmann model for black holes bears close resemblance to the 
multiperipheral model for bound state formation in the familiar partonic picture
of quantum field theory in the infinite momentum frame [11,12]. At lowest order, 
neglecting their interactions, the Matrix theory partons can be assumed to be 
in random Brownian motion filling the transverse $d-1$ dimensional spatial volume. 
This transverse space has been identified with the \lq\lq stretched horizon" of a 
neutral black hole in M theory. The rescattering of partons is then determined by 
the famous $v^{4}/r^{D-4}$ potential of semiclassical Matrix theory, dimensionally 
reduced to $d+1$ space-time dimensions [12]. This viewpoint suggests that the neutral black 
hole is the {\it generic} bound state into which the Matrix theory partons condense 
at a phase transition point. While the analogy is suggestive, note that it does 
{\it not} take into account the influence of partonic interactions.

Consider the Boltzmann gas model for the collective dynamics of D0-branes
as suggested by [2-9]. We begin by pointing out that, quite generically, the 
thermodynamic relation $S \sim N$ can be linked to the onset of a phase transition 
in a gas of $N$ distinguishable particles.  Evidence for a possible transition is
suggested by the singular behavior in the partition function when the degeneracy of 
states is comparable to the Gibbs-Boltzmann energy factor. The entropy is then 
proportional to the energy, which in turn, for a gas of distinguishable particles, 
is proportional to the number of particles. Note that this result can only 
be applied to the D0brane gas when the influence of interactions on the statistical 
properties of the free ensemble is negligible. We will argue later that these conditions 
can in fact be met in the special case of four noncompact spatial dimensions.

More precisely, the partition function of a Boltzmann gas can be written as
$$
Z = \prod_{k} \sum_{n_k}
g(k)^{n_k} e^{-\beta n_k \epsilon(k) }
=\prod_{k} {1 \over {1- g(k) e^{-\beta \epsilon(k) } }} \eqno(5)
$$
where  the level of degeneracy is denoted by
$g(k)$ [12-14].
Note that the partition function diverges if
$g(k) \sim e^{\beta \epsilon(k)}$ indicating
the onset of a phase transition. Typically, most of the particles are
found in one of the quantum states at some critical
value $\beta_{c}$.
There exists a direct analogy between this phenomenon and
Bose-Einstein condensation, except for the important fact that in
the case at hand the
\lq\lq condensation" takes place in coordinate, rather than momentum space.
At the phase transition point the entropy of the condensate,
$ S = \beta E + \ln{Z} $,
is directly proportional to the energy, $ S = \beta_{c} E $.
Since $E \sim {N \over \beta}$ for a gas of distinguishable particles, we see 
that at the critical temperature we recover the scaling relation $S\sim N$.
This relation has in fact appeared in previous discussions of phase
transitions and black hole physics that were made independent of the
connection to M/string theory (see, for example, [16]).

The Boltzmann gas model can in fact be directly motivated from the partition 
function of the ${\cal{N}}=16$ supersymmetric U(N)
Yang-Mills quantum mechanics [1,17,18]
describing the dynamics of nonrelativistic D0-branes $ Z_{M} = Tr \exp(-\beta H_{M})$,
with the Matrix theory Hamiltonian [1]
$$
H_{M} = R tr ({1 \over 2} P_a^2 + {1 \over 4} [X_a,X_b]^2 + \bar{\theta}
\gamma^{a}[X_a,\theta]). \eqno(6)
$$
In the limit of low velocities and large separations, a one-loop
evaluation of the Matrix theory partition function [18] gives the result
$$
Z_{M}= \exp \left ( -\beta \sum_{i=1}^{N} {{v_{i}^{2}} \over 2R}
+ \beta {15 \over 16} {{G_{11}} \over {R^{3}}}
\sum_{i < j}^{N} \int_{-{\beta}/2}^{{\beta}/2} dt
{{v_{ij}^4} \over (b_{ij}^{2} + v_{ij}^{2} t^{2})^{7/2}} \right ) \quad .
\eqno(7)
$$
The leading behavior of (7) captures an essential feature of
the multiperipheral model [10] for partonic dynamics; the
leading term in the partition function is of Maxwell-Boltzmann form, 
with $ E \sim {N \over \beta}$.
This immediately implies that at the phase transition point
where $S \sim \beta_{c} E$, we have the scaling relation $ S \sim N$. In this 
correspondence, the partition function of a gas of $N$ noninteracting 
distinguishable Boltzmann particles occupying spatial volume $V$ reads 
$ Z_{Boltzmann} \sim V^N {\beta}^{{{N(1-d)} \over {2}}} $ [3,13-15].

To summarize, there is evidence both from the partition function of a generic 
Boltzmann gas of distinguishable particles, and from the one--loop evaluation
of the partition function of Matrix theory, that is suggestive of a phase 
transition to a neutral black hole at $S \sim N$. However, there are pathologies 
in the specific heat of the Boltzmann gas which suggest a breakdown of the
assumptions behind the model in $d<6$ noncompact spatial dimensions. We will 
now show that these facts are simply explained by the results of an analysis 
of self-gravitating random walks due to Horowitz and Polchinski [10].

The Boltzmann gas model for black holes consists of D0-branes executing random walks 
in a classical background potential arising from long-range gravitational interactions.
The particles are confined to a box size, $R_S$, determined by the boost in the 
longitudinal direction [2,3]. The statistical mechanics of random walks with 
a long-range attractive interaction is precisely that studied by Horowitz and 
Polchinski [10], in conjunction with the statistical correspondence between long 
excited string states and black holes [18-23]. The question addressed by these 
authors is {\it whether} a long excited string can, in fact, be confined to a 
region of order the string scale, $R_S \sim \sqrt{\alpha '}$, prior to entering 
the domain of strong string coupling. The equality of entropies (and masses) 
at the phase transition, $S_{bh} \sim S_{string}$, requires: 
$$
{{R_{S}^{d-1}} \over G_{D}} \sim  \sqrt{N_s}
 \quad .  \eqno(8)
$$
Using the usual relation between string coupling and the $d+1$-dimensional
Newton's constant, $G_D = g^2 ({\sqrt \alpha '})^{d-1}$, the critical value of the string coupling 
at which the transition could occur is determined by the string oscillator level, 
$g_c \sim N_s^{1/4}$. Horowitz and Polchinski have shown that the critical coupling 
at which self-interactions become non-negligible is
$$
g_o \sim N_s^{{(d-6)}\over{6}} \quad , \eqno(9)
$$
which is {\it larger} than $g_c$ in all dimensions $d$$>$$4$.

Consider the implications of this result. For $d$$=$$3$ noncompact spatial dimensions, 
where the entropy of the Boltzmann gas was infinite, we have $g_0$$<$$g_c$. The gravitational 
self-interactions must be taken into account {\it before} the onset of the phase transition. 
In the special case of $d$$=$$4$ dimensions, the interactions become non-negligible 
precisely when the string is confined to a region of order the string scale,
$g_c $$=$$g_o$, and the phase transition to a black hole occurs within the domain of 
weak string coupling. In $d$$=$$5$ eqn.(9) suggests the interesting possibility of 
{\it hysteresis}, the string continues to shrink due to the gravitational 
self-interactions past $g$$\sim$$g_c$ until the critical coupling $g_0$$\sim$$N_s^{-1/6}$
when it collapses into a black hole. In the reverse transition, the black
hole phase remains a good description until $g$$\sim$$g_c$. Finally, for all $d$$>$$6$
the transition is in the strong coupling regime and the black hole does not have a 
weakly coupled string-like phase.

These results exactly match the dependence on spacetime dimensionality found in the 
scaling analysis of the Boltzmann model, with opposing domains of validity, and with 
$d$$=$$6$ noncompact spatial dimensions the borderline case. In the special case of
$d$$=$$4$, we can give an independent argument for a phase transition in a dilute gas 
of D0-branes reminiscent of that for free highly excited strings. Our result suggests 
that, in four noncompact spatial dimensions, a long Hagedorn string has the same 
thermodynamic properties as a condensate of $N$$=$$N_s$ distinguishable D0-branes. 
This condensate is in the $N$$>$$S$ regime of the Boltzmann gas.

Recall the random walk picture of highly excited strings. The density of states function 
for free excited strings displays an exponential growth in the number of states. For
excited states at oscillator level $N_s$, we have [22-26], 
$$
d(E) \sim E^{-p} \exp( - \beta E) \quad , \eqno(10)
$$
where $E \sim \sqrt {N_s}$ and $p$$>$$1$. A
snapshot of the string at any instant of time resembles a Brownian motion such
that the average relative separation of any two string \lq\lq bits" is a point
executing motions in a volume $E^{d/2}$ [22]. Thus, in the absence of interactions, 
the size of the random walk is $ l \sim N_s^{1/4} {\alpha '}^{1/2}$. The
microcanonical ensemble corresponding to such random motions is to be identified
with the motions of a {\it dilute} gas of $N$ weakly interacting D0-branes in
$d$ noncompact spatial dimensions. From the analysis of [10], we expect the
effect of interactions to be negligible in the special case of self-gravitating
random walks in $d$$=$$4$.

The absence of interactions at the phase transition in the $d$$=$$4$ dilute gas 
of D0-branes suggests the following simplification in the computation of the 
asymptotic density of states. We can use an observation due to Vafa [26] (see 
also the computation in [27][28]) on the density of ground states of $m$ D0-branes, 
delocalized in the four directions in the world-volume of a D4-brane {\it transverse} 
to their common 0+1 dimensional intersection. The generating functional for the 
ground state degeneracies of the $m0$$-$$4$ bound state are easily obtained by 
noting the isomorphism to the partition function of a conformal field theory of 
four free bosons and four free fermions. (This is also the world sheet content 
of the holomorphic part of a type II string in light cone gauge [26].) The generating 
functional $d(m)$ takes the form (eq. (112) of [28])
$$
\sum_{m=0}^{\infty} q^{m} d(m) = 256
\prod_{k=1}^{\infty}({{1+q^{k}} \over {1-q^{k}}})^{8} \quad  , \eqno(11)
$$
and upon setting $m=N$ with $N$ large, we obtain the asymptotic degeneracy 
$d(N) \sim \exp^{2 \pi {\sqrt{2N}} } $, consistent with a condensate of entropy 
$S = \ln{d(N)} \sim {\sqrt{N}}$. This is evidence of a phase transition in
the density of states analogous to the Hagedorn transition of free strings
with $N$$\sim $$N_s$. 

Notice that this is in disagreement with the scaling relations of the Boltzmann
model that follow from the naive rule of thumb $S$$\sim$$N$. Equating the 
condensate mass of $N$ D0-branes to the mass of a Hagedorn string of size the
string scale, would imply the scaling formula $ M^{2} \sim {{N^2} \over {R_S }}$. 
At $R_S$ of order the string scale, this gives $M^{2} \sim {{N_s} \over {\alpha'}}$,
which would imply $N^{2} \sim N_s$, in disagreement with the result above.

Our result seems to suggest that it may be natural to regard Hagedorn strings
as bound states of D0-branes, although we have only been able to provide 
concrete evidence in the special case of $d$$=$$4$ dimensions. If one examines 
the microcanonical ensemble of free strings in the Hagedorn limit, as in [22], 
one finds that, to first order,
$S \sim \beta_{H} E$. But this relation is what we expect from the general 
argument given above (eq. (3)),
if we identify $\beta_{c}$ with $\beta_{H}$ at the Hagedorn point.
But we know that according to Matrix theory black holes are
simply bound states of D0-branes. Also in the Hagedorn limit
the entropy is proportional to energy [25], just as in the Boltzmann
model of Matrix black holes [2-9]. The interpretation of Hagedorn strings 
as bound states of D0-branes is natural from the point of view of Matrix string 
theory [29].

Indeed, according to [29] the large $N$ limit of a two dimensional
${\cal{N}}=8$ supersymmetric Yang-Mills field theory describes
the non-perturbative dynamics of ten-dimensional light cone
string theory. D0-branes correspond to states with non-zero
electric flux in the supersymmetric Yang-Mills theory, the flux being
equal to the D0-brane
charge. The sector of the Yang-Mills theory that accounts for states which do
not carry electric flux describes strings in the background of D0-branes.
Moreover, the strongly coupled limit of this two dimensional
${\cal{N}}=8$ supersymmetric Yang-Mills theory, which corresponds
to the weakly coupled limit of string theory, is
conjectured to be an
${\cal{N}}=8$ superconformal field theory with the central charge $8N$ [29].
Using this fact and Cardy's formula for the density of states
in two-dimensional conformal field theory [30] we estimate
the entropy to be $S\sim N$, which is again consistent with the Boltzmann
model of collective D0-brane dynamics. Thus, from the viewpoint of Matrix 
String Theory, it is natural to regard Hagedorn strings as bound states of 
D0-branes that form at a {\it particular} value of the Yang-Mills coupling.

The resulting picture meshes nicely with the well known entropy vs 
energy interpretation of the Hagedorn transition [25].
The entropy of long strings wins over energy close to the
Hagedorn transition point, indicating the increase in the
density of states and the importance of string interactions [24].
The entropy of a sufficiently excited Hagedorn string is proportional 
to its length, $\sqrt N_s$. And, from the correspondence above, this is
simply the number of partonic constituents $N$ of a Matrix theory 
\lq\lq string", i.e., the number of string 
\lq\lq bits" in Susskind's picture [19].  
It has been observed in the past that the high temperature limit of 
D-dimensional string theory behaves like a field theory in {\it two} 
space-time dimensions [24], indicating a drastic reduction 
in the number of degrees of freedom as compared to ordinary field theory.
This observation follows from a T-duality in the time direction, as
applied to the free energy of noninteracting strings. Note that this 
observation is compatible with the description of Hagedorn strings 
provided by Matrix string theory [29]: two-dimensional ${\cal{N}}=8$ 
supersymmetric $U(N)$ Yang-Mills quantum field theory in the strong
coupling regime, for which $S \sim N$.

\vskip .1in
{\bf Acknowledgements}

It is a pleasure to thank Kirill Krasnov, Miao Li, and Joe Polchinski
for interesting discussions. We would like to thank J\"{u}rgen Baacke for 
bringing reference [31] to our attention.


\vskip.1in
{\bf References}
\vskip .1in
\item{1.} T. Banks, W. Fischler, S. H. Shenker and L. Susskind, Phys.
Rev. D55 (1997) 5112. For a review and further references, see T.Banks,
hep-th/9706168; D. Bigatti and L. Susskind, hep-th/9712072.
\item{2.} T. Banks, W. Fischler, I. Klebanov and L. Susskind,
hep-th/9709091.
\item{3.} T. Banks, W. Fischler, I. Klebanov and L. Susskind,
 hep-th/9711005.
\item{4.} I. Klebanov and L. Susskind, hep-th/9709108.
\item{5.} E. Halyo, hep-th/9709225; S. Das, S. Mathur, S. Kalyana Rama and
P. Ramadevi, hep-th/9711003.
\item{6.} G. Horowitz and E. Martinec, hep-th/9710217.
\item{7.} M. Li, hep-th/9710226.
\item{8.} M. Li and E. Martinec, hep-th/9801070.
\item{9.} T. Banks, W. Fischler and I. Klebanov, hep-th/9712236.
\item{10.} G. Horowitz and J. Polchinski, hep-th/9707170.
\item{11.} J. Kogut and L. Susskind, Phys. Rep. 8 (1973) 78.
\item{12.} H. Awata, S. Chaudhuri, M. Li, and D. Minic, 
Phys. Rev. {\bf D57} (1998), hep-th/9706083.
\item{13.} H. Liu and A. Tseytlin, hep-th/9712063; D. Minic, hep-th/9712202.
\item{14.} A. Strominger, Phys. Rev. Lett. 71 (1993) 3397; I. V. Volovich,
hep-th/9608137.
\item{15.} J. Hartle, R. Stolt, and J. Taylor, Phys. Rev. D2 (1970) 1759;
O. W. Greenberg, Phys. Rev. Lett. 64 (1990) 705; Phys. Rev. D43 (1991) 4111.
\item{16.} K. Krasnov, gr-qc/9603025; gr-qc/9605047 and references
therein.
\item{17.} M. Baake, P. Reinicke, V. Rittenberg, J. Math. Phys. 26 (1985)
1070; R. Flume, Ann. Phys. 164 (1985) 189; C. Claudson and
M. Halpern, Nucl. Phys. B250 (1985) 689.
\item{18.} N. Ohta and Jian-Ge Zhou, hep-th/9801023; O. Tafjord and
V.Periwal, hep-th/9711046.
\item{19.} L. Susskind, hep-th/9309145.
\item{20.} See, for example, A. Peet, hep-th/9712253, and references within; 
J. Maldacena, hep-th/9607235.
\item{21.} G. Horowitz and J. Polchinski, hep-th/9612146.
\item{22.} P. Salomonson and B.-S. Skagerstam, Nucl. Phys. B268
(1986) 349; Physica A158 (1989) 499 and references within; D. Mitchell and 
N. Turok, Phys. Rev. Lett. 58 (1987) 1577; Nucl. Phys. B294 (1987) 1138.
\item{23.} B. Sathiapalan, Phys. Rev. D35 (1987) 3227; I. A. Kogan, JETP Lett. 45
(1987) 709.  
\item{24.} J. J. Atick and E. Witten, Nucl. Phys. B310 (1988), 291, and
references therein.
\item{25.} R. Brandenberger and C. Vafa, Nucl. Phys. B316 (1988) 391.
\item{26.} C. Vafa, Nucl. Phys. B463 (1996) 415. See, also, 
Section 6 of C. Vafa and E. Witten, Nucl. Phys. B431 (1994) 3.
\item{27.} M. Douglas, D. Kabat, P. Pouliot, and S. Shenker, hep-th/9608024. 
\item{28.} J. Polchinski, {\it TASI Lectures on D-branes}, pp. 44,
 hep-th/9611050.
\item{29.} R. Dijkgraaf, E. Verlinde and H. Verlinde, hep-th/9703030.
\item{30.} J. Cardy, Nucl. Phys. B270 (1986) 186.
\item{31.} J. Baacke, Acta Phys. Polonica {\bf B8} 625 (1977).
\end